\documentclass{PoS}
\usepackage{epsfig,subfigure,color,natbib}

\definecolor{purple}{rgb}{1,0,1}

\newcommand{\thetacut}{\theta_{\rm cut}}
\newcommand{\lcut}{\ell_{\rm cut}}
\newcommand{\lmax}{\ell_{\rm max}}

\newcommand{\tgpu}{t_{\tt ARKCoS}}
\newcommand{\tcpu}{t_{\tt libpsht}}

\bibpunct[; ]{(}{)}{;}{a}{}{,}

\title{Accelerating convolutions on the sphere with hybrid GPU/CPU kernel splitting}

\ShortTitle{Accelerating convolutions on the sphere}

\author{\speaker{P.M. Sutter}\\
Department of Physics, University of Illinois at Urbana-Champaign, Urbana, IL 61801, USA\\
 UPMC Univ Paris 06, UMR7095, Institut d'Astrophysique de Paris, F-75014, Paris, France \\
CNRS, UMR7095, Institut d'Astrophysique de Paris, F-75014, Paris, France \\
Center for Cosmology and Astro-Particle Physics, Ohio State University, Columbus, OH, USA\\
        E-mail: \email{sutter@iap.fr}}

\author{Benjamin D. Wandelt\\
 UPMC Univ Paris 06, UMR7095, Institut d'Astrophysique de Paris, F-75014, Paris, France \\
CNRS, UMR7095, Institut d'Astrophysique de Paris, F-75014, Paris, France \\
Department of Physics, University of Illinois at Urbana-Champaign, Urbana, IL 61801, USA
Department of Astronomy, University of Illinois at Urbana-Champaign, Urbana, IL 61801, USA
       }

\author{Franz Elsner \\
 UPMC Univ Paris 06, UMR7095, Institut d'Astrophysique de Paris, F-75014, Paris, France \\
CNRS, UMR7095, Institut d'Astrophysique de Paris, F-75014, Paris, France \\
       }

\abstract{We present a general method for accelerating by more than an order of magnitude
the convolution of pixelated function on the sphere with a radially-symmetric kernel.
Our method  splits  the kernel into a compact real-space, and a compact spherical harmonic space component
that can then be convolved in parallel using an inexpensive commodity GPU and a CPU, 
respectively. We provide models for the computational cost of both real-space and Fourier space convolutions and an estimate for the approximation error.
Using these models we can determine the 
optimum split that minimizes the wall clock time for the convolution while satisfying the desired error bounds.
We apply this technique to the problem of simulating a cosmic microwave 
background sky map at the resolution typical of the high resolution maps 
of the cosmic microwave background anisotropies produced by the 
Planck space craft. For the main Planck CMB science channels we 
achieve a speedup of over a factor of ten, assuming an acceptable fractional rms error of order $10^{-5}$ in the (power spectrum of the) output map.
}

\FullConference{Big Bang, Big Data, Big Computers\\
                 September 19-21, 2012\\
                 Laboratoire Astroparticule et Cosmologie, 10 rue A. Domon et L. Duquet, 75205 Paris 13, France}

\begin{document}
%-------------------------------------------------------------------------------
----------------------------------------------------------------------------
\section{Introduction}
\label{sec:introduction}

Cosmic microwave background (CMB) experiments, 
such as Planck~\citep{PlanckCollaboration2011},
the Atacama Cosmology Telescope~\citep{KOSOWSKY2003},
the South Pole Telescope~\citep{Ruhl2004},
and CMBPol~\citep{Baumann2009}
promise a great wealth of cosmological 
and astrophysical information~\citep{Smoot2010}.
The most common operation in CMB data analysis consists of convolving  a real or a synthetic map with a radial kernel.
Large numbers of such  smoothing or filtering operations are necessary for 
many critical data analysis applications, such as the simulation of CMB maps~\citep{Gorski2005},
map-making from multichannel maps~\citep{Tegmark1997, Natoli2001, Stompor2001, Patanchon2008, Sutton2010}, iterative calculation of inverse covariance weighted data, e.g. in the context of optimal power spectrum estimation or Wiener filtering~\citep{Wandelt2004}
wavelet analysis~\citep{Hobson1999, Martinez-Gonzalez2002, Vielva2004}, 
point-source removal~\citep{Tegmark1998, Gonzalez-Nuevo2006},
and analysis of errors.
The future Euclid mission~\citep{Laureijs2011} will resolve the 
sky to sub-arcsecond resolution, and one technique for identifying 
overdensities in such a map is via convolution with a filter.

Until recently, the near-exclusive practice in the CMB community to compute radial kernel convolutions was to use the spherical convolution theorem: transformation to spherical harmonic space, multiplying the spherical harmonic coefficients with the l-space representation of the radial kernel and back-transformation to pixel space. As a consequence of the ubiquity of radial kernel convolution for data analysis on the sphere and the ready availability of software implementing the discrete forward and backward fast Spherical Harmonic Transformation (SHTs), this has become the major application for SHTs. Interest in the actual $a_{lm}$ is relatively rare by comparison.

Graphics Processing Units (GPUs) offer a promising solution 
to the  computational challenges posed by radial kernel convolution 
to current and upcoming 
data sets on the sphere~\citep{Brunner2007, Barsdell2010, Fluke2011}  due to their low cost and high degree of parallelism.
Indeed, the recent rise of 
cheap GPU hardware and associated extensive programming 
libraries have led to their use in
many applications in astrophysics, such as
the analysis of the Lyman-$\alpha$ forest~\citep{Greig2011},
dust temperature calculations~\citep{Jonsson2010},
magnetohydrodynamics~\citep{Pang2010},
adaptive-mesh refinement simulations~\citep{Schive2010},
analysis of data from the Murchison Widefield Array~\citep{Wayth2007},
volume renderings of spectral data from the Australian 
Square Kilometer Array Pathfinder mission~\citep{Hassan2011},
and visualizations of large-scale data sets~\citep{Szalay2008}.

While GPU implementations of the SHT \citep{HUPCAIoanO.2010,Szydlarski2011} have only achieved modest 
speed-ups,~\citet[][hereafter EW11]{Elsner2011} 
tackled the problem of spherical 
convolutions for compact radial kernels by specifically designing an algorithm  adapted to benefit from high degree of parallelism and memory bandwidth for compact kernels.
Compared to the serial time of a highly optimized implementation of the Fast SHT algorithm, EW11 demonstrated a speed-up 
of up to a factor of 
60 using a commodity GPU costing \$500 with the further benefit of strongly suppressing Fourier ringing artifacts.  Other approaches, 
such as optimizing traditional algorithms~\citep{Muciaccia1997}
and using large-scale computing resources~\citep{GhellerC.2007}, either 
do not scale as efficiently or do not exploit readily available hardware.

The main limitation of the method described by EW11 is that significant speed-ups can only be achieved for relatively compact kernels. 
While there are still many applications for such compact kernels,
truncating non-compact kernels to realize large performance improvements
can lead to unreasonable artefacts in the resulting convolved maps.
To take advantage of the power of GPUs with kernels of arbitrary size, 
we must split the given kernel between a real-space portion, which will 
be applied using a GPU, and an $\ell$-space (i.e., Fourier) 
portion, which will be applied 
using traditional CPU methods. Each portion of the full kernel will then 
necessarily be truncated, resulting in a small, but predictable, error. 
Given an upper bound for an acceptable error, we must determine the 
optimal splitting between real- and $\ell$-space in order to achieve 
maximum performance.

In this work we present 
{\tt scytale} \footnote{We take the name from the ancient cryptographic 
system where only rods of a precise radius could be used to decode messages.}, 
a tool for splitting kernels between truncated real- and $\ell$-space 
portions, estimating the errors due to the truncations,
 and discovering the optimum truncations for a given kernel.
We apply this tool to determine the expected speedup 
when splitting a given kernel 
between the GPU code {\tt ARKCoS} of EW11 and the CPU code 
{\tt libpsht} of~\citet{Reinecke2011}.
In Section~\ref{sec:optimization} we discuss our strategy for 
splitting kernels, estimating errors, and determining the optimum truncations. 
We present an analysis of the errors and our optimization results in 
Section~\ref{sec:results}, followed by a discussion and conclusion 
in Section~\ref{sec:conclusions}.

%-------------------------------------------------------------------------------
\section{Splitting formalism \& Optimization Strategy}
\label{sec:optimization}

We decompose a given kernel $K_\ell$ into truncated $\ell$-space and real-space 
portions, which we denote as $\widehat{K}_\ell$ and $\widehat{K}_\theta$, 
respectively. We may then construct an approximate kernel as
\begin{equation}
  \widetilde{K}_\ell = \widehat{K}_\ell + P_{\ell \theta} \widehat{K}_\theta,
\label{eq:approx}
\end{equation}
where $P_{\ell \theta}$ is a Legendre transformation operator.
We truncate the $\ell$-space kernel to a limit $\lcut$ and the 
real-space kernel to a limit $\thetacut$.
Once we have the truncated kernels, we can 
evaluate the resulting error by taking the fractional root mean square:
\begin{equation}
  \sigma^2 = \alpha^2 \frac{\sigma_{\rm rms}^2} 
                   {1 / 4 \pi \sum (2 \ell+1) K_\ell^2},
\label{eq:error}
\end{equation}
where the sum run from 0 to $\lmax$.
The constant $\alpha$ represents any additional errors introduced 
by the actual convolution, such as those caused by 
single-precision arithmetic and inadequate kernel interpolation, 
and must be empirically determined.
Thus, given a particular kernel, this procedure allows us to 
identify values of $\lcut$ and $\thetacut$ that 
satisfy a given error bound.

If a particular $\lcut$ and $\thetacut$ satisfy an error bound, we 
then estimate the computational cost associated with the truncated kernels.
We assume the real-space portion will be solved using {\tt ARKCoS}
on a GPU, so we denote the cost as $\tgpu$. Similarly, 
we assume the $\ell$-space kernel will be solved using the standard 
library {\tt libpsht} on a CPU, 
and hence we will denote the cost as $\tcpu$.
The cost for applying the real-space GPU kernel is
\begin{equation}
  \tgpu = 0.0232{\rm s}~\thetacut + 2.428{\rm s} 
  %\tgpu = 3~{\rm s}~\frac{\thetacut \lmax^2 \log{\lmax}}{60'~4096^2 \log{4096}}
\label{eq:costGPU}
\end{equation}
and the cost for the $\ell$-space CPU kernel is
\begin{equation}
  \tcpu = 160{\rm s}~\frac{\lcut^2 \lmax}{4096^3}.
\label{eq:costCPU}
\end{equation}
Above, $\thetacut$ is in arcminutes.
To determine these scalings we used
an NVIDIA GeForce GTX 480 GPU and a 2.8 GHz Intel Core2 Quad CPU.
Our GPU scaling is different than the study of EW11 due to updated 
NVIDIA drivers.
Note that the CPU timing assumes the use of only a single core.
We assume throughout a data set with 
HEALPix~\citep{Gorski2005} 
resolution $n_{\rm side} = 2048$ and $\lmax=4096$, consistent 
with Planck observations~\citep{MennellaA.2011}.
Furthermore, we assume a power spectrum derived from WMAP 7-year 
results~\citep{Komatsu2011}.

We assume that the GPU and CPU portions can be solved in 
parallel, and hence our goal for a given kernel is to find the 
pair $(\lcut, \thetacut)$ that satisfies the error bound and 
at which $\tgpu=\tcpu$, minimizing the overall cost.

%-------------------------------------------------------------------------------
\section{Results}
\label{sec:results}

We study radially-symmetric kernels of the type
\begin{equation}
  K_\ell = \sqrt{C_\ell} B_\ell,
\label{eq:kernel}
\end{equation}
where $C_\ell$ is the expected power in the 
given $\ell$-space bin and $B_\ell$ is the 
Legendre transform of a beam. We assume an identical band limit of $\lmax$
for both the input power spectrum and the kernel. 
These particular kernels have a wide variety 
of applications. We assume a Gaussian beam 
with a given FWHM. For this analysis, we will also assume 
$C_\ell^{\rm input} \sim 1$ (that is, the case of simulating 
CMB maps with uncorrelated noise).

\begin{figure} 
  \centering 
  {\includegraphics[width=\columnwidth]{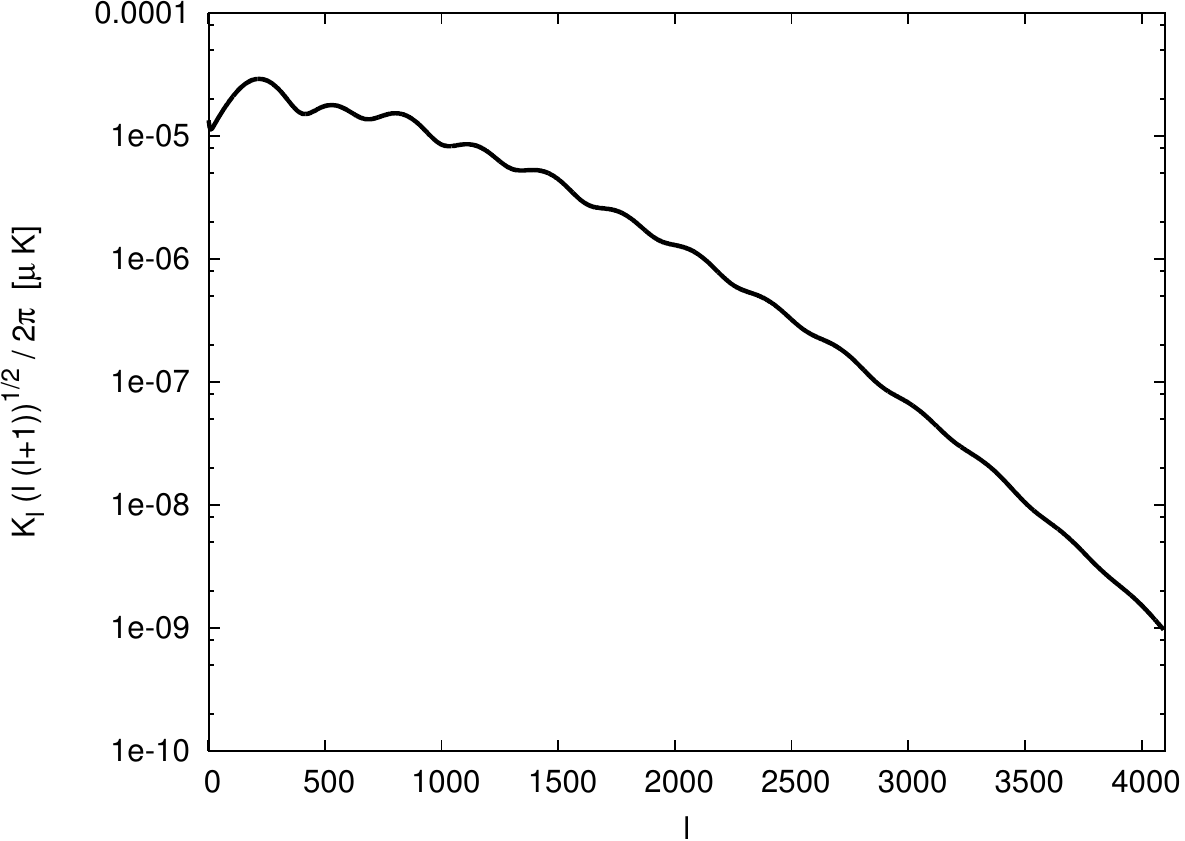}}
  \caption{Example kernel 
           for a beam with 7 arcmin FWHM.}
\label{fig:kernel}
\end{figure}

We begin with an analysis of splitting a single kernel. 
We show in Figure~\ref{fig:kernel} an example kernel produced 
with a 7 arcmin FWHM beam. We truncate the kernel and the input 
power spectrum at $\lmax=4096$. This narrow beam produces wide 
support to significantly high $\ell$: only past $\ell \approx 2000$ 
does the kernel drop below $1 \%$ of $\sqrt{C_\ell}$.

After splitting, the $\ell$-space kernel faithfully reproduces the low-$\ell$ 
portion of the full kernel while the real-space kernel matches the 
high-$\ell$ regime. In order to fit the behavior of the full kernel 
at high $\ell$, the real-space kernel produces large oscillations at 
low $\ell$, which are compensated by percent-level adjustments in 
the $\ell$-space kernel. Summed together, these kernels reproduce the 
full input kernel, except at the very highest $\ell$ where the low 
magnitudes make a full fit difficult.

Even though our computational approach 
damps oscillations in $\ell$-space (where the fits to the 
full input kernel take place) we see rapid oscillations in the actual 
kernel that {\tt ARKCoS} uses in its real-space approach.
We must accurately 
interpolate this kernel, especially at small angles, 
in the convolution algorithm in order to both 
recover the high-$\ell$ behavior and correctly calculate 
the systematic offsets 
present in the low-$\ell$ portion of the approximate kernel.
To do this, we employ a simple bias where we place half the available 
interpolation nodes within the first $1/16$th of the available support; 
in this case, within 7.5 arcmin. We found this bias to be a good 
compromise between the need to carefully interpolate the innermost 
portions of the kernel and the need to maintain a sufficient 
number of interpolation points throughout the rest of the kernel.  

\begin{figure} 
  \centering 
  {\includegraphics[width=\columnwidth]{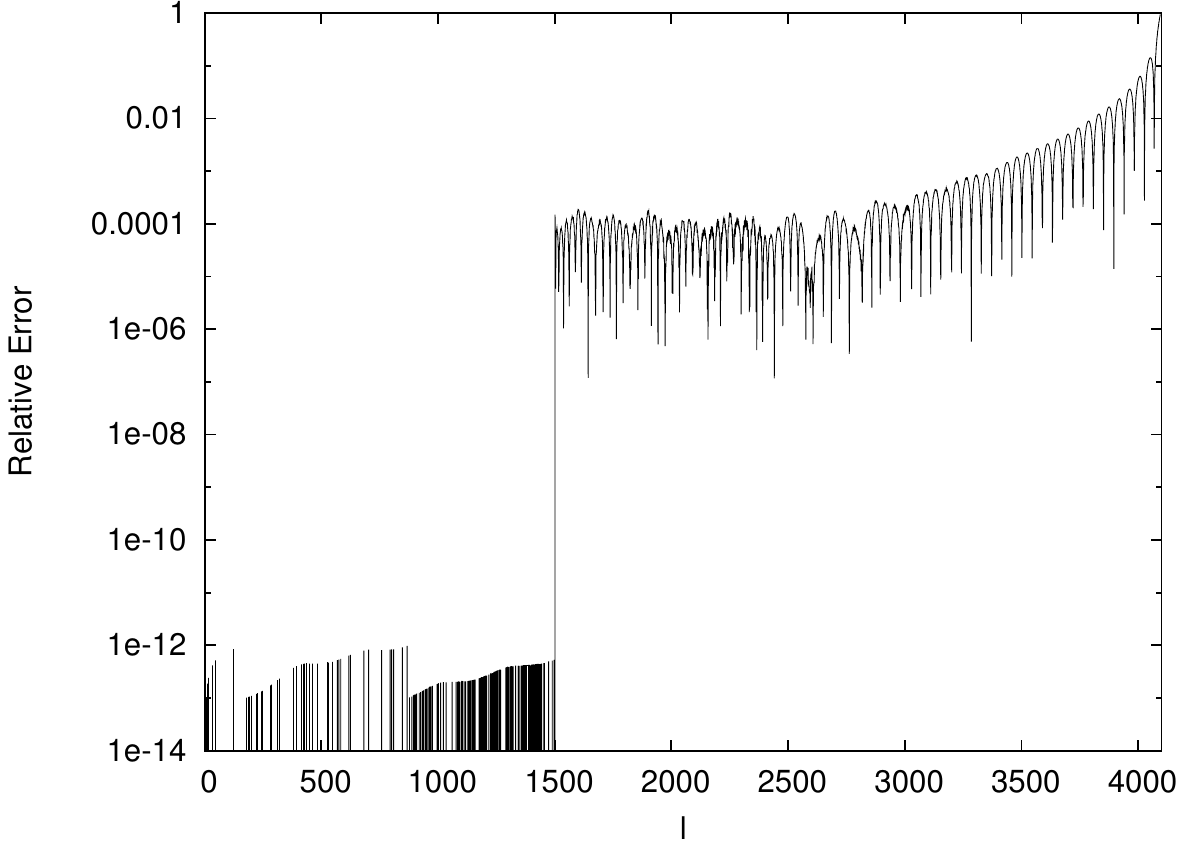}}
  \caption{Estimated relative error of the 
           example approximate kernel $\widetilde{K}_\ell$ to 
           the full kernel $K_\ell$. Shown is the relative error as 
           a function of $\ell$.
           For this example, the 
           $\ell$-space kernel is truncated to $\lcut=1500$ and the
           real-space kernel to $\thetacut=240$ arcmin. }
\label{fig:relerror}
\end{figure}

The approximate kernel faithfully represents the full input kernel 
below the truncation threshold of the $\ell$-space kernel at 
$\ell=1500$, which we see in Figure~\ref{fig:relerror}. 
In this figure we show the relative error, defined as
\begin{equation}
  \sigma_\ell = \log_{10} \left| 1 - 
                 \frac{\widetilde{K}_\ell}{K_\ell} \right|.
\label{eq:relerror}
\end{equation}
In this figure we see three distinct regimes. The first, from $\ell=$0-1500 
where the $\ell$-space kernel dominates, has essentially zero error.
From $\ell=$1500 to roughly 3000, we maintain a relative error of roughly 
$10^{-5}$. In this region the real-space kernel is best able to 
reproduce the full input kernel. Finally, at the highest $\ell$, the 
real-space kernel has difficulty following the input kernel and the errors 
begin to exponentially diverge, reaching $100 \%$ 
relative error at $\lmax=4096$. However,
the beam strongly suppresses the kernel here and the high-magnitude 
low-$\ell$ portion
dominates our error estimate. 
Therefore we can ultimately satisfy a given overall error bound.

\begin{figure} 
  \centering 
  {\includegraphics[width=\columnwidth]{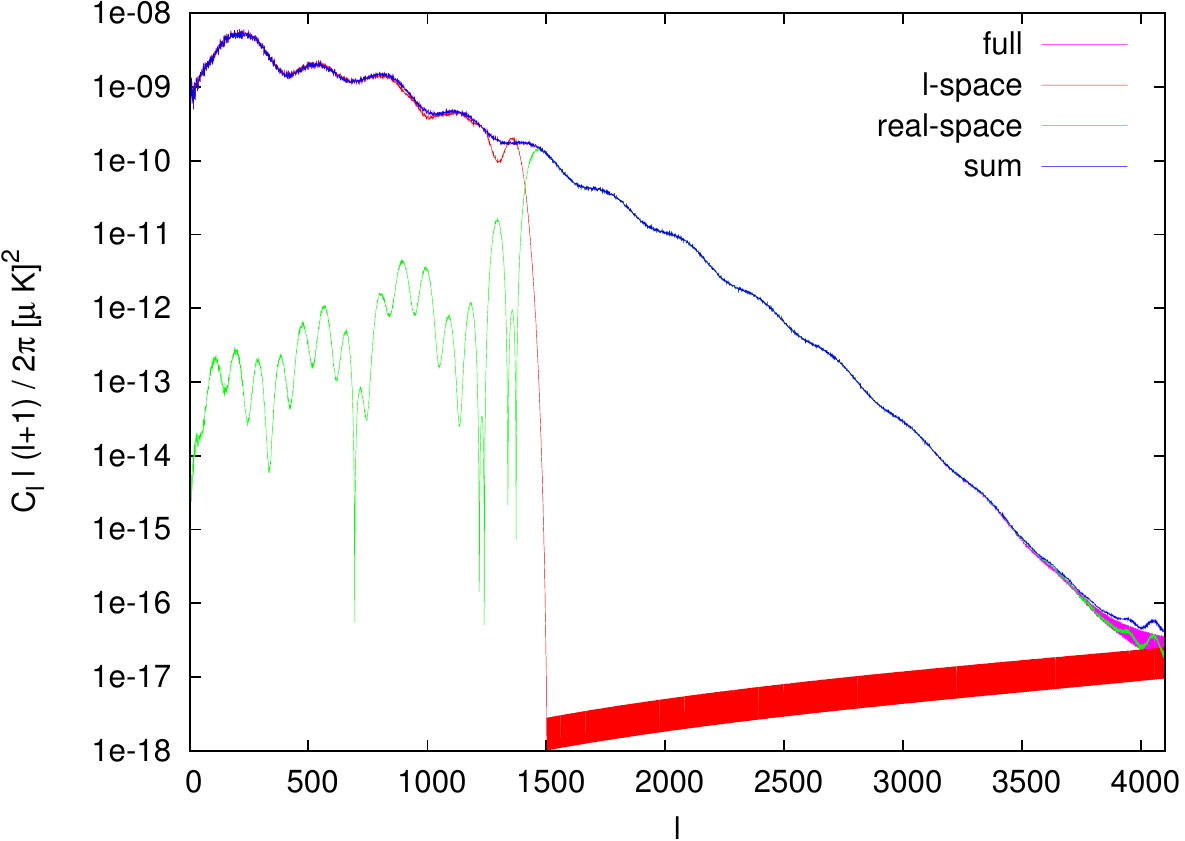}}
  \caption{Derived power spectra after convolving a uniform-noise map
           with various kernels.
           The kernels used are: 
           the full $\ell$-space kernel $K_\ell$ (pink),
           the truncated $\ell$-space kernel $\widehat{K}_\ell$ (red),
           and the truncated real-space kernel $\widehat{K}_\theta$ (green).
           The blue line shows the power spectrum of the map
           created by summing the individual maps of 
           the two truncated kernels.
           For this example, 
           the $\ell$-space kernel is truncated to $\lcut=1500$ and the
           real-space kernel to $\thetacut=240$ arcmin. 
}
\label{fig:spectra}
\end{figure}

To evaluate the actual performance of each kernel, we applied them to 
a uniform-noise input map and extracted the spectra.
We compare these spectra in Figure~\ref{fig:spectra}. We show the 
power spectrum after convolving with the full $\ell$-space kernel $K_\ell$,
the truncated $\ell$-space kernel $\widehat{K}_\ell$,
and the truncated real-space kernel $\widehat{K}_\theta$. We also show 
the power spectrum of the summed map. We see that we are able to recover 
the desired power spectrum using the truncated kernels, except at the 
highest $\ell$ range, where interpolation errors and the limitations of 
single-precision arithmetic in the GPU introduce deviations.

\begin{figure} 
  \centering 
  {\includegraphics[width=\columnwidth]{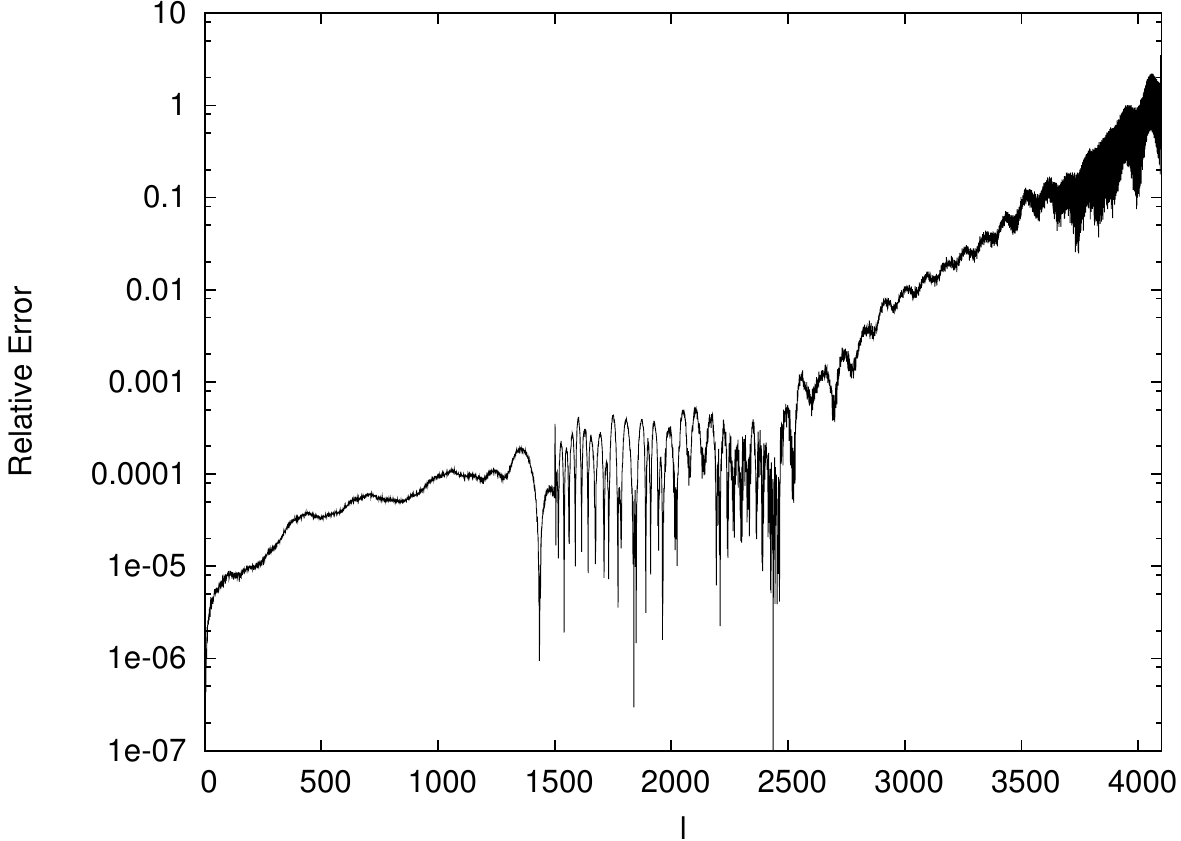}}
  \caption{Actual relative error of the 
           approximate kernel $\widetilde{K}_\ell$ to 
           the full kernel $K_\ell$ after convolution. 
           Shown is the relative error as 
           a function of $\ell$.
           For this example, 
           the $\ell$-space kernel is truncated to $\lcut=1500$ and the
           real-space kernel to $\thetacut=240$ arcmin. 
          }
\label{fig:spectrumerror}
\end{figure}

Figure~\ref{fig:spectrumerror} shows the relative error between the 
power spectrum obtained by summing the maps produced by the truncated 
kernels and spectrum obtained by using the full $\ell$-space kernel.
We see similar structure to the estimated relative error, but in this 
case the errors are not negligible below $\lcut=1500$. Here, the difficulty 
of adding the small component due to the real-space kernel to the 
$\ell$-space kernel is apparent. After $\ell=1500$ we see 
small oscillations around the full power spectrum followed by the 
expected exponential rise in the relative error. Altogether, 
we found the total error to be a factor of five higher than estimated 
due to these numerical effects. 
Thus we set the constant $\alpha$ in Eq.(\ref{eq:error}) to five.

\begin{figure*} 
  \centering 
  \subfigure[full kernel]{
    \includegraphics[width=\columnwidth]{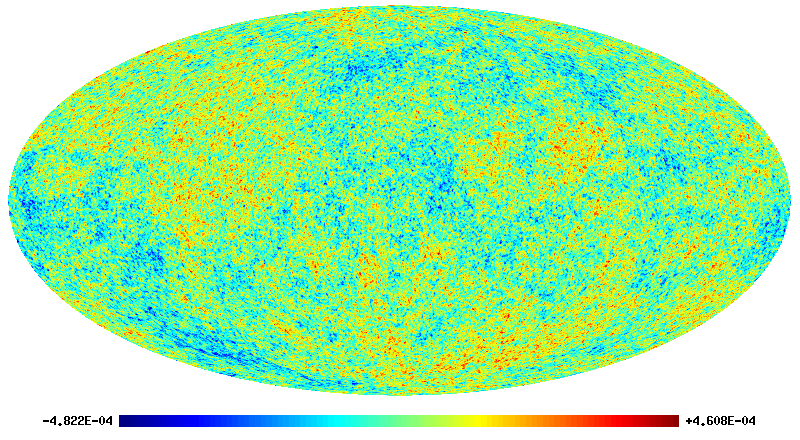}
  }
  \subfigure[difference]{
    \includegraphics[width=\columnwidth]{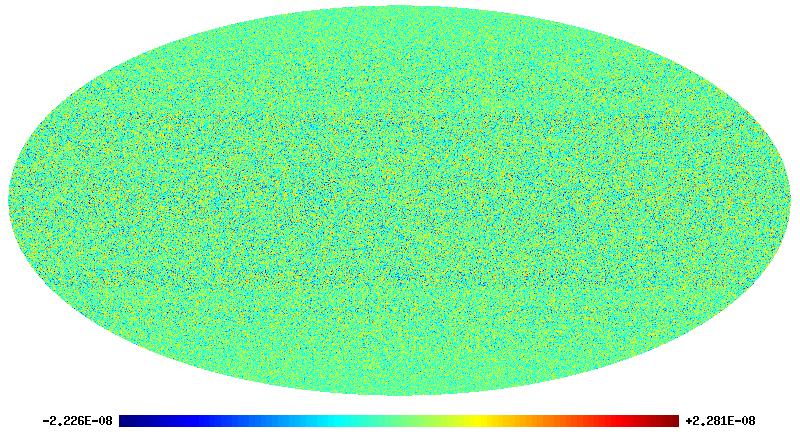}
  }
  \caption{(a) Map after convolving a uniform-noise input map with 
           the full $\ell$-space kernel $K_\ell$.
           (b) The difference between the map
           in panel (a) and the map constructed by summing 
           the convolution outputs of  
           the truncated $\ell$-space kernel $\widehat{K}_\ell$
           and the truncated real-space kernel $\widehat{K}_\theta$.
           For this example, 
           the $\ell$-space kernel is truncated to $\lcut=1500$ and the
           real-space kernel to $\thetacut=240$ arcmin.}
\label{fig:maps}
\end{figure*}

\begin{figure*} 
  \centering 
  \subfigure[full kernel]{
    \includegraphics[width=0.31\textwidth]{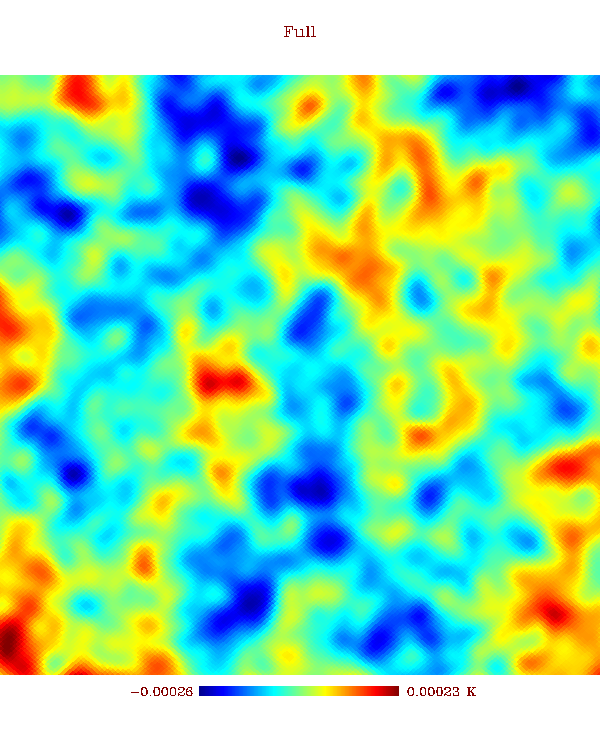}
  }
  \subfigure[l-space difference]{
    \includegraphics[width=0.31\textwidth]{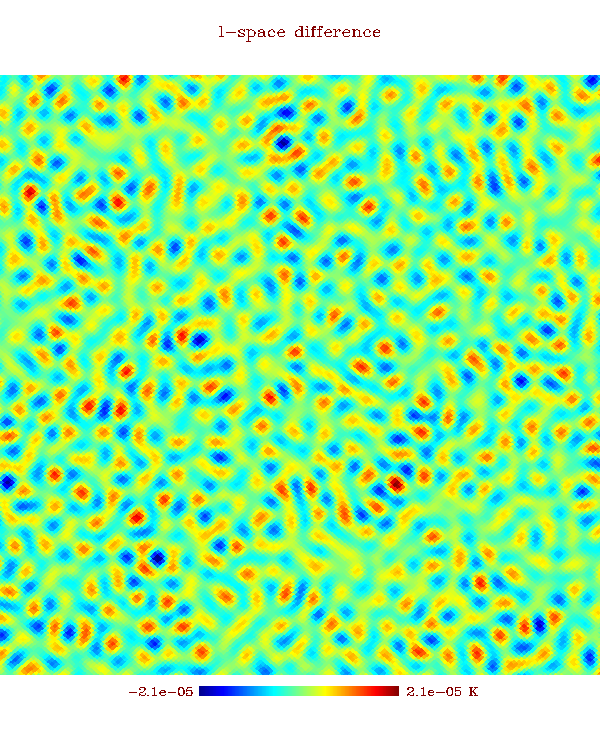}
  }
  \subfigure[real-space kernel]{
    \includegraphics[width=0.31\textwidth]{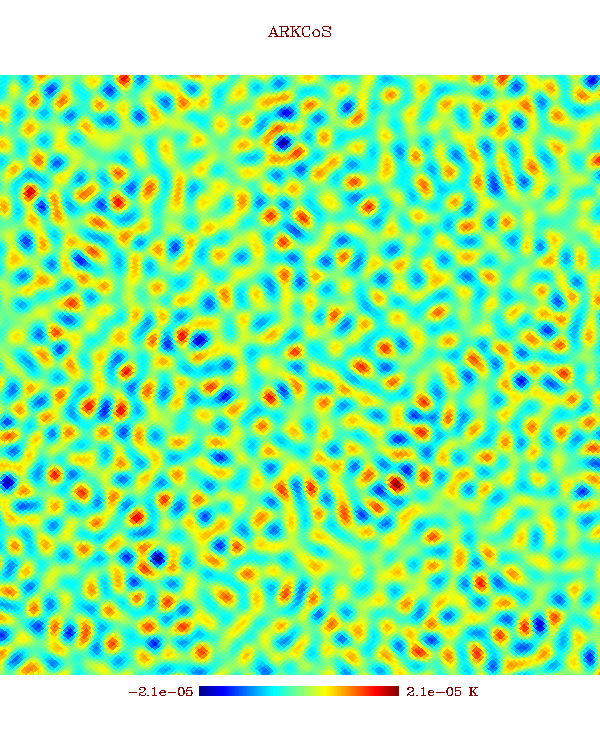}
  }
  \caption{(a) Five-degree patch of the map convolved with the full 
               unsplit kernel. 
           (b) Difference between the map in panel (a) and the 
               map produced by convolving with
               the truncated $\ell$-space kernel $\widehat{K}_\ell$.
           (c) Map created by convolving with 
               the truncated real-space kernel $\widehat{K}_\theta$.
           For this example, 
           the $\ell$-space kernel is truncated to $\lcut=1500$ and the
           real-space kernel to $\thetacut=240$ arcmin.}
\label{fig:smallmaps}
\end{figure*}

In Figure~\ref{fig:maps} we show the map after convolving with the 
full $\ell$-space kernel. We also show the difference between this 
map and sum of the maps produced by convolution with the truncated 
$\ell$-space and real-space kernels. We maintain small errors throughout 
the entire map, with the largest errors at the smallest scales, 
as expected.
In Figure~\ref{fig:smallmaps} we show a 5-degree patch of the same 
maps. We see that the 
$\ell$-space kernel reproduces the full map to percent-level accuracy.
However, the real-space kernel is necessary to correctly construct the 
small-scale power and reduce the error to acceptable limits. 

\begin{figure} 
  \centering 
  {\includegraphics[width=\columnwidth]{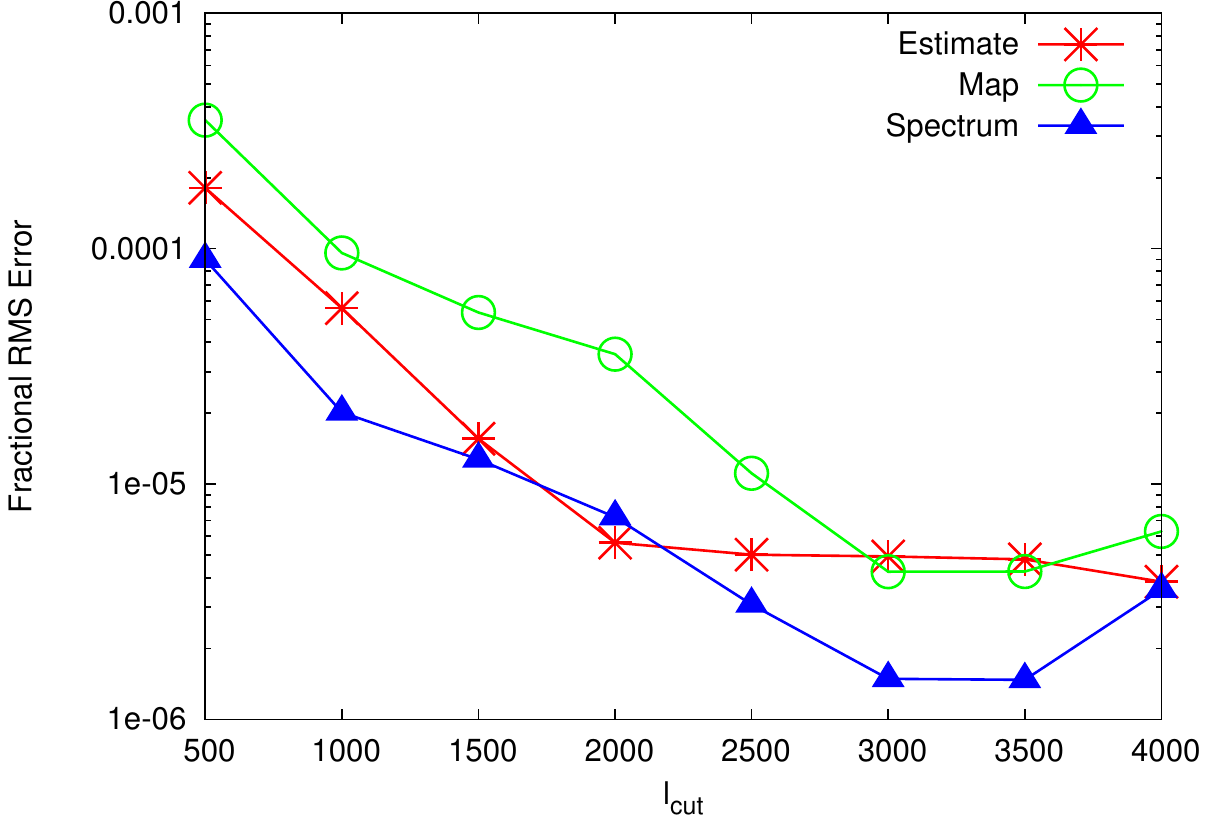}}
  \caption{Estimated 
           RMS error computed by {\tt scytale} 
           with
           $\alpha=5$ (red stars)
           versus actual RMS error in the maps produced by convolution 
           with a uniform-noise map (green circles) and the RMS error 
           in the power spectra 
           derived from those maps (blue triangles). The lines do not 
           represent data but are shown as visual aids.}
\label{fig:error}
\end{figure}

We compare our estimated RMS error to the actual 
map and power spectra errors 
in Figure~\ref{fig:error} for a selection of $\lcut$ values with a
fixed $\thetacut=240$ arcmin and the same 7 arcmin beam that we have 
thus far used. For this plot, we have set the empirically-determined 
constant $\alpha$ to five. With this chosen constant, our error 
estimate matches the actual error in the power spectra until 
an $\lcut$ of 2500. At higher $\lcut$ values, we overestimate the 
spectrum errors, but since this lies below our chosen 
error bound of $10^{-5}$ (see below) we choose to maintain this 
value of $\alpha$. The maps tend to produce higher errors, but since 
our quantity of interest is the derived power spectrum, we choose to 
match those errors.

With all this in place
we now turn to our scanning strategy and results of our optimization 
study. We examine beams with 5-10 arcmin FWHM, which are most relevant 
to the Planck mission~\citep{MennellaA.2011}. Table~\ref{tab:cuts} shows 
the optimum $(\lcut, \thetacut)$ pairs for five of the ten beam sizes 
studied, assuming a maximum error bound of $10^{-5}$.
Below 6 arcmin we could not find suitable truncations that still 
maintained our desired error bound.
We see that all truncations are essentially identical, indicating that 
the ability to split these kernels is binary: either no optimum 
truncations can be found, but if optimum truncations can be 
found they will be very aggressive. 
For these beam sizes, the optimum $\lcut$ values that 
satisfy the error bounds are significantly below $\lmax$, which promise 
significant enhancements in performance.

\begin{table}
\centering
\caption{Optimum $\lcut$ and $\thetacut$ pairs for each beam FWHM 
         studied, assuming an error bound of $10^{-5}$. }
\begin{tabular}{ccc}
\hline
\hline
Beam FWHM (arcmin) & $\ell_{\rm cut}$ & $\theta_{\rm cut}$ (arcmin) \\
\hline
 7 & 1158 & 390 \\
 8 & 1070 & 390 \\
 9 & 1055 & 360 \\
10 & 979 & 360 \\
11 & 1014 & 330 \\
12 & 960 & 330 \\
13 & 940 & 330 \\
14 & 929 & 300 \\
15 & 961 & 270 \\
\hline
\end{tabular}
\label{tab:cuts}
\end{table}

\begin{figure} 
  \centering 
  {\includegraphics[width=\columnwidth]{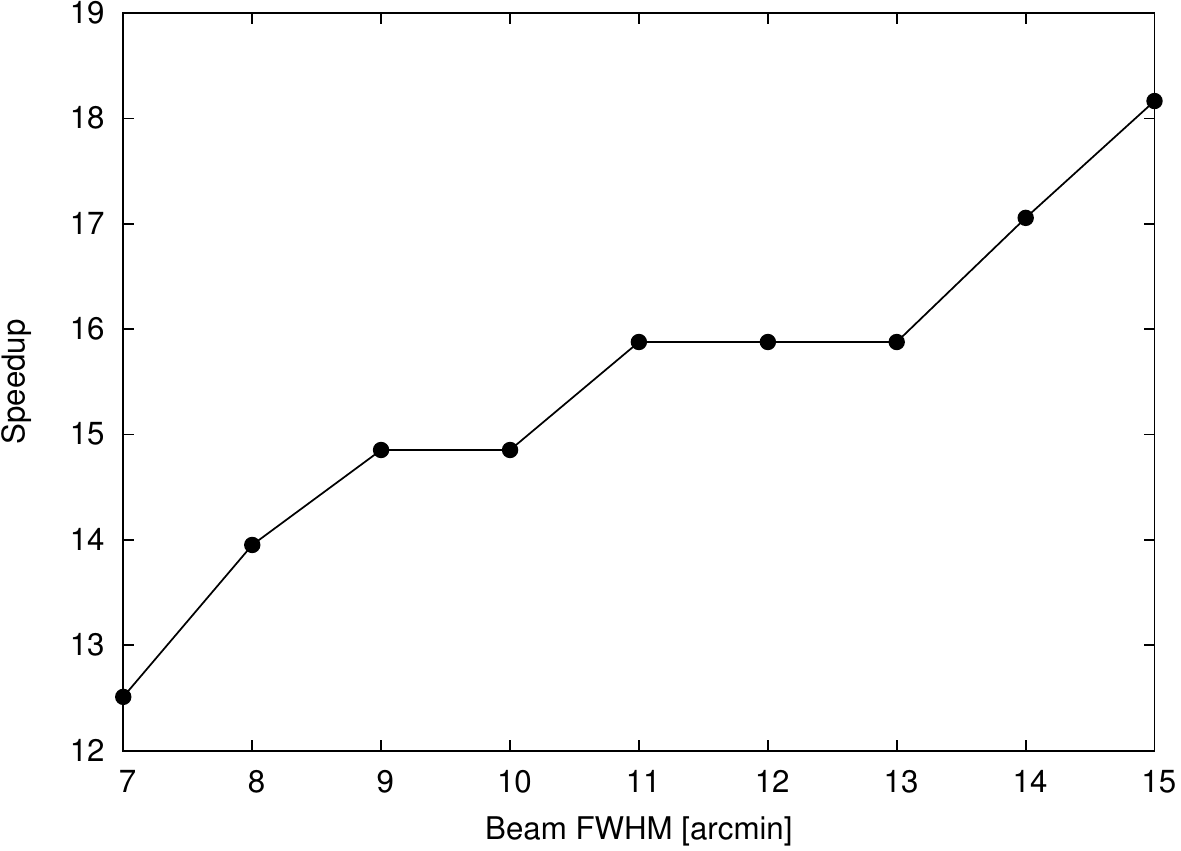}}
  \caption{Speedup versus beam FWHM assuming an overall error bound 
           of $10^{-5}$. }
\label{fig:speedup}
\end{figure}

We show in Figure~\ref{fig:speedup} the speedup versus beam FWHM for these 
beam sizes and our error bound of $10^{-5}$. 
We define the scaling as the time to solution 
with our split approach relative to the cost of applying the
entire kernel (i.e., up to $\lmax$) on the 
CPU with {\tt libpsht}. Below 7 arcmin, we find no optimum truncations 
and hence do not show them. We see significant performance gains
 above 7 arcmin, with the speedups plateauing in the range 12-15. This 
speedup implies a reduction in the computational time from 160 seconds to 
approximately 12 seconds for a single convolution operation. Since all the 
truncations are essentially the same above 7 arcmin, we find nearly 
identical speedups regardless of the beam size.

%-------------------------------------------------------------------------------
\section{Conclusions}
\label{sec:conclusions}

We have introduced and 
discussed a method for splitting radially-symmetric kernels into 
truncated real- and Fourier-space components and estimating the errors 
associated with such splitting. 
We have validated our error estimation 
by performing convolutions with the truncated kernels and computing 
the actual resulting error. 
We have found that for Planck-sized
data sets, a large range of kernels can be split into 
significantly truncated portions while still maintaining an acceptable 
($\sim 10^{-5}$) error bound, leading to significant speedups.

Our analysis was focused on an ideal case; i.e., situations where 
there is no noise and where the input power spectrum remains 
flat. This is the worst-case scenario.
In the case where noise dominates the high-$\ell$ regime we found 
speedups of order $\sim 20$, since we could relax the criterion of 
strictly matching the structure of the full kernel in this region.  

Our approach is currently limited to $\ell \sim 4000$ due to the 
finite amount of fixed memory available on single current-generation GPUs. 
An all-sky convolution up to $\ell=8000$ or $16000$ would require 
splitting the problem across multiple GPUs, as discussed below. However, current experiments that probe 
this regime, such as ACT~\citep{KOSOWSKY2003} and 
SPT~\citep{Ruhl2004}, only map on the order of hundreds of square degrees.
By re-orienting their survey maps onto the polar cap, we can keep the 
number HEALPix rings small and exploit our algorithm with currently-available 
GPUs.

The {\tt ARKCoS} code also has a CPU-based implementation, allowing 
our approach to work on homogeneous architectures. While the speedups 
in the CPU-only case are not as significant, we can still take advantage of the 
parallelism offered by the compact real-space kernels. In this 
scenario, the truncated $\ell$-space kernel can be convolved 
using traditional parallel spherical harmonic transform operations 
on a few cores 
(such as {\tt ccSHT} \footnote{http://crd-legacy.lbl.gov/~cmc/ccSHTlib/doc/index.html}), 
where the parallel scalability is strongest, while 
the truncated real-space kernel can be convolved using many cores 
in parallel in the manner described above. 
 
Kernel splitting enables the efficient 
allocation of resources for tackling large data sets; in our case, by 
applying real-space kernels with a GPU and 
$\ell$-space kernels with a CPU. 
We have applied this kernel splitting scheme to an optimization study to 
find the realistic speedups associated with splitting a kernel between 
a compact portion to be solved on a GPU and the remainder on a CPU. 
Applying this to kernels and data sets appropriate for the Planck mission, 
we find that this splitting technique can lead to over a factor of 
ten speedup compared to traditional fully CPU-based approaches. 
This significantly improves the feasibility of many necessary and 
important data analysis operations, such as wavelet analysis, point 
source removal, and map making. 

%-------------------------------------------------------------------------------
%-------------------------------------------------------------------------------
\section*{Acknowledgments}

The authors acknowledge support from NSF Grants AST-0908902, 
AST-0908693 ARRA, and AST-0708849.
This material also is based upon work supported in part by NSF Grant AST-1066293 and the hospitality of the Aspen Center for Physics.

\bibliography{big3}		
\bibliographystyle{apj}	\nocite{*}

\end{document}